\input harvmac
\def\newdate{7/31/2003} 

\def\a{\alpha} 
\def\b{\beta} \def\g{\gamma} \def\l{\lambda} \def\d{\delta} \def\e{\epsilon} \def\t{\theta}  
\def\k{\kappa}
    
  \def\p{\partial} \def\half{{1\over 2}}

\def\z{{\bar z}} 
\def\m{\mu} 
\def\n{\nu}
 

  
\lref\lrp{  
U.~Lindstr\"om, M.~Ro\v cek, and P.~van Nieuwenhuizen, in preparation.   
}  
  
\lref\pr{  
P. van Nieuwenhuizen, in {\it Supergravity `81},   
Proceedings First School on Supergravity, Cambridge University Press, 1982, page 165.  
}  
  
\lref\polc{  
J.~Polchinski,  
{\it String Theory. Vol. 1: An Introduction To The Bosonic String,}  
{\it String Theory. Vol. 2: Superstring Theory And Beyond,}  
{\it  Cambridge, UK: Univ. Pr. (1998) 531 p}.  
}  
  
\lref\superstring{  
M.~B.~Green and  J.~H.~Schwarz, {\it Covariant Description Of Superstrings,}   
Phys.\  Lett.\ {\bf B 136} (1984) 367; M.~B.~Green and J.~H.~Schwarz,  
  {\it Properties Of The Covariant Formulation Of Superstring Theories,}  
  Nucl.\ Phys.\ {\bf B 243} (1984) 285\semi  
M.~B.~Green and C.~M.~Hull, QMC/PH/89-7  
{\it Presented at Texas A and M Mtg. on String Theory, College  
  Station, TX, Mar 13-18, 1989}\semi  
R.~Kallosh and M.~Rakhmanov, Phys.\ Lett.\  {\bf B 209} (1988) 233\semi  
U. ~Lindstr\"om, M.~Ro\v cek, W.~Siegel,   
P.~van Nieuwenhuizen and A.~E.~van de Ven, Phys. Lett. {\bf B 224} (1989)   
285, Phys. Lett. {\bf B 227}(1989) 87, and Phys. Lett. {\bf B 228}(1989) 53;   
S.~J.~Gates, M.~T.~Grisaru, U.~Lindstr\"om, M.~Ro\v cek, W.~Siegel,   
P.~van Nieuwenhuizen and A.~E.~van de Ven,  
{\it Lorentz Covariant Quantization Of The Heterotic Superstring,}  
Phys.\ Lett.\  {\bf B 225} (1989) 44;   
A.~Mikovic, M.~Rocek, W.~Siegel, P.~van Nieuwenhuizen, J.~Yamron and  
A.~E.~van de Ven, Phys.\ Lett.\  {\bf B 235} (1990) 106;   
U.~Lindstr\"om, M.~Ro\v cek, W.~Siegel, P.~van Nieuwenhuizen and  
A.~E.~van de Ven,   
{\it Construction Of The Covariantly Quantized Heterotic Superstring,}  
Nucl.\ Phys.\  {\bf B 330} (1990) 19 \semi  
F. Bastianelli, G. W. Delius and E. Laenen, Phys. \ Lett. \ {\bf  
  B229}, 223 (1989)\semi  
R.~Kallosh, Nucl.\ Phys.\ Proc.\ Suppl.\  {\bf 18B}  
  (1990) 180 \semi  
M.~B.~Green and C.~M.~Hull, Mod.\ Phys.\ Lett.\  {\bf A5} (1990) 1399\semi   
M.~B.~Green and C.~M.~Hull, Nucl.\ Phys.\  {\bf B 344} (1990) 115\semi  
F.~Essler, E.~Laenen, W.~Siegel and J.~P.~Yamron, Phys.\ Lett.\  {\bf B 254} (1991) 411\semi   
  F.~Essler, M.~Hatsuda, E.~Laenen, W.~Siegel, J.~P.~Yamron, T.~Kimura  
  and A.~Mikovic,   
  Nucl.\ Phys.\  { \bf B 364} (1991) 67\semi   
J.~L.~Vazquez-Bello,  
  Int.\ J.\ Mod.\ Phys.\  {\bf A7} (1992) 4583\semi  
E. Bergshoeff, R. Kallosh and A. Van Proeyen, ``Superparticle  
  actions and gauge fixings'', Class.\ Quant.\ Grav {\bf 9}   
  (1992) 321\semi  
C.~M.~Hull and J.~Vazquez-Bello, Nucl.\ Phys.\  {\bf B 416}, (1994) 173 [hep-th/9308022]\semi  
P.~A.~Grassi, G.~Policastro and M.~Porrati,  
{\it Covariant quantization of the Brink-Schwarz superparticle,}  
Nucl.\ Phys.\ B {\bf 606}, 380 (2001)  
[hep-th/0009239].  
}  
  
 \lref\big{
M.~Henneaux,
{\it Brst Cohomology Of The Fermionic String,}
Phys.\ Lett.\ B {\bf 183}, 59 (1987); W.~Siegel,
{\it Boundary Conditions In First Quantization,}
Int.\ J.\ Mod.\ Phys.\ A {\bf 6}, 3997 (1991); 
N.~Berkovits, M.~T.~Hatsuda and W.~Siegel,
{\it The Big picture}
Nucl.\ Phys.\ B {\bf 371}, 434 (1992)
[hep-th/9108021].
}
  
\lref\bv{  
N. Berkovits and C. Vafa,  
{\it $N=4$ Topological Strings}, Nucl. Phys. B {\bf 433}, 123 (1995),   
hep-th/9407190.}  
  
\lref\fourreview{N. Berkovits,  {\it Covariant Quantization Of  
The Green-Schwarz Superstring In A Calabi-Yau Background,}  
Nucl. Phys. {\bf B 431} (1994) 258, ``A New Description Of The Superstring,''  
Jorge Swieca Summer School 1995, p. 490, hep-th/9604123.}  
  
\lref\OoguriPS{  
H.~Ooguri, J.~Rahmfeld, H.~Robins and J.~Tannenhauser,  
{\it Holography in superspace,}  
JHEP {\bf 0007}, 045 (2000)  
[hep-th/0007104].  
}  
  
\lref\bvw{  
N.~Berkovits, C.~Vafa and E.~Witten,  
{\it Conformal field theory of AdS background with Ramond-Ramond flux,}  
JHEP {\bf 9903}, 018 (1999)  
[hep-th/9902098].  
}  

\lref\wittwi{  
E.~Witten,  
{\it An Interpretation Of Classical Yang-Mills Theory,}  
Phys.\ Lett.\ B {\bf 77}, 394 (1978);   
E.~Witten,  
{\it Twistor - Like Transform In Ten-Dimensions,}  
Nucl.\ Phys.\ B {\bf 266}, 245 (1986)}  
 
\lref\SYM{  
W.~Siegel,  
{\it Superfields In Higher Dimensional Space-Time,}  
Phys.\ Lett.\ B {\bf 80}, 220 (1979)\semi  
B.~E.~Nilsson,  
{\it Pure Spinors As Auxiliary Fields In The Ten-Dimensional   
Supersymmetric Yang-Mills Theory,}  
Class.\ Quant.\ Grav.\  {\bf 3}, L41 (1986);   
B.~E.~Nilsson,  
{\it Off-Shell Fields For The Ten-Dimensional Supersymmetric   
Yang-Mills Theory,} GOTEBORG-81-6\semi  
S.~J.~Gates and S.~Vashakidze,  
{\it On D = 10, N=1 Supersymmetry, Superspace Geometry And Superstring Effects,}  
Nucl.\ Phys.\ B {\bf 291}, 172 (1987)\semi  
M.~Cederwall, B.~E.~Nilsson and D.~Tsimpis,  
{\it The structure of maximally supersymmetric Yang-Mills theory:    
Constraining higher-order corrections,}  
JHEP {\bf 0106}, 034 (2001)  
[hep-th/0102009];   
M.~Cederwall, B.~E.~Nilsson and D.~Tsimpis,  
{\it D = 10 superYang-Mills at O(alpha**2),}  
JHEP {\bf 0107}, 042 (2001)  
[hep-th/0104236].  
}  
\lref\har{  
J.~P.~Harnad and S.~Shnider,  
{\it Constraints And Field Equations For Ten-Dimensional   
Super-Yang-Mills Theory,}  
Commun.\ Math.\ Phys.\  {\bf 106}, 183 (1986).  
}  
\lref\wie{  
P.~B. Wiegman,  
{\it Multivalued Functionals And Geometrical Approach   
For Quantization Of Relativistic Particles And Strings,}   
Nucl.\ Phys.\ B {\bf 323}, 311 (1989).  
}  
\lref\purespinors{\'E. Cartan, {\it Lecons sur la th\'eorie des spineurs},   
Hermann, Paris (1937)\semi  
C. Chevalley, {\it The algebraic theory of Spinors},   
Columbia Univ. Press., New York\semi  
 R. Penrose and W. Rindler,   
{\it Spinors and Space-Time}, Cambridge Univ. Press, Cambridge (1984)   
\semi  
P. Budinich and A. Trautman, {\it The spinorial chessboard}, Springer,   
New York (1989).  
}  
\lref\coset{  
P.~Furlan and R.~Raczka,  
{\it Nonlinear Spinor Representations,}  
J.\ Math.\ Phys.\  {\bf 26}, 3021 (1985)\semi  
A.~S.~Galperin, P.~S.~Howe and K.~S.~Stelle,  
{\it The Superparticle and the Lorentz group,}  
Nucl.\ Phys.\ B {\bf 368}, 248 (1992)  
[hep-th/9201020].  
}  
  
\lref\GS{M.B. Green, J.H. Schwarz, and E. Witten, {\it Superstring Theory,}   
 vol. 1, chapter 5 (Cambridge U. Press, 1987).    
}  
\lref\carlip{S. Carlip,   
{\it Heterotic String Path Integrals with the Green-Schwarz   
Covariant Action}, Nucl. Phys. B284 (1987) 365 \semi R. Kallosh,   
{\it Quantization of Green-Schwarz Superstring}, Phys. Lett. B195 (1987) 369.}   
 \lref\john{G. Gilbert and   
D. Johnston, {\it Equivalence of the Kallosh and Carlip Quantizations   
of the Green-Schwarz Action for the Heterotic String}, Phys. Lett. B205   
(1988) 273.}   
\lref\sok{E. Sokatchev, {\it   
Harmonic Superparticle}, Class. Quant. Grav. 4 (1987) 237\semi   
E.R. Nissimov and S.J. Pacheva, {\it Manifestly Super-Poincar\'e   
Covariant Quantization of the Green-Schwarz Superstring},   
Phys. Lett. B202 (1988) 325\semi   
R. Kallosh and M. Rakhmanov, {\it Covariant Quantization of the   
Green-Schwarz Superstring}, Phys. Lett. B209 (1988) 233.}    

\lref\many{S.J. Gates Jr, M.T. Grisaru,   
U. Lindstrom, M. Rocek, W. Siegel, P. van Nieuwenhuizen and   
A.E. van de Ven, {\it Lorentz-Covariant Quantization of the Heterotic   
Superstring}, Phys. Lett. B225 (1989) 44\semi   
R.E. Kallosh, {\it Covariant Quantization of Type IIA,B   
Green-Schwarz Superstring}, Phys. Lett. B225 (1989) 49\semi   
M.B. Green and C.M. Hull, {\it Covariant Quantum Mechanics of the   
Superstring}, Phys. Lett. B225 (1989) 57.}    

 \lref\fms{D. Friedan, E. Martinec and S. Shenker,   
{\it Conformal Invariance, Supersymmetry and String Theory},   
Nucl. Phys. B271 (1986) 93.}  

\lref\kawai{  
T.~Kawai,  
{\it Remarks On A Class Of BRST Operators,}  
Phys.\ Lett.\ B {\bf 168}, 355 (1986).}  

 \lref\ufive{N. Berkovits, {\it   
Quantization of the Superstring with Manifest U(5) Super-Poincar\'e   
Invariance}, Phys. Lett. B457 (1999) 94, hep-th/9902099.}    
 
\lref\berkovitsFE{  
N.~Berkovits,  
{\it Super-Poincar\'e covariant quantization of the superstring,}  
JHEP { 0004}, 018 (2000).}

\lref\berkovitsPH{  
N.~Berkovits and B.~C.~Vallilo,  
{\it Consistency of super-Poincar\'e covariant superstring tree amplitudes,}  
JHEP { 0007}, 015 (2000)  
[hep-th/0004171].}

\lref\berkovitsNN{  
N.~Berkovits,  
{\it Cohomology in the pure spinor formalism for the superstring,}  
JHEP { 0009}, 046 (2000)  
[hep-th/0006003].}

\lref\berkovitsWM{  
N.~Berkovits,  
{\it Covariant quantization of the superstring,}  
Int.\ J.\ Mod.\ Phys.\ A { 16}, 801 (2001)  
[hep-th/0008145].}

\lref\berkovitsYR{  
N.~Berkovits and O.~Chandia,  
{\it Superstring vertex operators in an AdS(5) x S(5) background,}  
Nucl.\ Phys.\ B {\bf 596}, 185 (2001)  
[hep-th/0009168].}

\lref\berkovitsMX{  
N.~Berkovits and O.~Chandia,  
{\it Lorentz invariance of the pure spinor BRST cohomology for the  superstring,}  
[hep-th/0105149].}

\lref\berkovitsRB{ 
N.~Berkovits,   
{\it Covariant quantization of the superparticle using pure spinors,} 
[hep-th/0105050].    
}  
 
 \lref\Membrane{  
N.~Berkovits,  
{\it Towards covariant quantization of the supermembrane,}  
[hep-th/0201151].  
}

\lref\berkovitsUS{  
N.~Berkovits,  
{\it Relating the RNS and pure spinor formalisms for the superstring,}  
[hep-th/0104247].
}

\def\berko{\berkovitsFE, \berkovitsPH, \berkovitsNN, \berkovitsWM, \berkovitsYR, 
\berkovitsMX, \berkovitsRB, \berkovitsUS}

\def\berkox{[7 - 14]\rlap{\phantom{\berko}}}


\lref\Grassione{  
P.~A.~Grassi, G.~Policastro, M.~Porrati and P.~van Nieuwenhuizen,  
{\it Covariant quantization of superstrings without pure spinor constraints},   
JHEP {\bf 10} (2002) 054, 
[hep-th/0112162].}
  
\lref\Grassitwo{  
P.~A.~Grassi, G.~Policastro, and P.~van Nieuwenhuizen,  
{\it The massless spectrum of covariant superstrings},   
JHEP {\bf 11} (2002) 004, 
[hep-th/0202123].}

\lref\Grassithree{
P.~A.~Grassi, G.~Policastro and P.~van Nieuwenhuizen,
{\it On the BRST cohomology of superstrings with / without pure spinors,}
[hep-th/0206216].}

\lref\Grassifour{
P.~A.~Grassi, G.~Policastro and P.~van Nieuwenhuizen,
{\it The covariant quantum superstring and superparticle from their  classical actions,}
Phys. Lett. {\bf B 553} (2003) 96, 
[hep-th/0209026].
}

\lref\Grassifive{
P.~A.~Grassi, G.~Policastro and P.~van Nieuwenhuizen,
{\it Yang-Mills theory as an illustration of the covariant quantization of  superstrings}, 
Proceedings 3$^{rd}$ Sacharov Conference, 2002 Moscow, 
[hep-th/0211095].}

\lref\Grassisix{
P.~A.~Grassi, G.~Policastro and P.~van Nieuwenhuizen,
{\it Introduction to the covariant quantization of  superstrings}, 
Proceedings Leuven Conference, 2002, Leuven, [hep-th/0302147].
}  
 
 \lref\Grassiseven{P.A. Grassi, G. Policastro and P. van Nieuwenhuizen, 
 {\it  Superconformal Algebras and Quantum Superstrings}, to appear.
 }

\def\grassi{\Grassione, \Grassitwo, \Grassithree, \Grassifour, \Grassifive, \Grassisix}

\def\grassix{[1 - 6]\rlap{\phantom{\grassi}}}
\def\twosix{[2,6]}

 \lref\figue{
J.~M.~Figueroa-O'Farrill and S.~Stanciu,
{\it Nonreductive WZW models and their CFTs, II: N=1 and N=2 cosets,}
Nucl.\ Phys.\ B {\bf 484}, 583 (1997)
[hep-th/9605111]; 
J.~M.~Figueroa-O'Farrill and S.~Stanciu,
{\it N=1 and N=2 cosets from gauged supersymmetric WZW models,}
[hep-th/9511229]; 
.~M.~Figueroa-O'Farrill and S.~Stanciu,
{\it Nonreductive WZW models and their CFTs,}
Nucl.\ Phys.\ B {\bf 458}, 137 (1996)
[hep-th/9506151].
}

\lref\figueB{
J. Figueroa-O'Farrill, {\it N=2 Structure in String Theories} 
J. Math. Phys. {\bf 38}(11) (1997), 5559
}

\lref\grr{
M.B. Green, {\it Supertranslations, Superstrings and Chern-Simons Forms}, 
Phys. Lett. B {\bf 223} (1989) 157}

\lref\csm{W.~Siegel, {\it Classical Superstring Mechanics}, Nucl. Phys. {\bf B 263} (1986) 93\semi   
W.~Siegel, {\it Randomizing the Superstring}, Phys. Rev. D {\bf 50} (1994) 2799.  
}     
%
 
\lref\WittenZZ{  
E.~Witten,  
{\it Mirror manifolds and topological field theory,}  
hep-th/9112056.  
}  
  
\lref\wichen{  
E.~Witten,  
{\it Chern-Simons gauge theory as a string theory,}  
[hep-th/9207094].  
}  
  
\lref\howe{P.S. Howe, {\it Pure Spinor Lines in Superspace and   
Ten-Dimensional Supersymmetric Theories},   
Phys. Lett. B258 (1991) 141, Addendum-ibid.B259 (1991) 511\semi   
P.S. Howe, {\it Pure Spinors, Function Superspaces and Supergravity   
Theories in Ten Dimensions and Eleven Dimensions}, Phys. Lett. B273 (1991)   
90.}  
  
\lref\tonin{ 
I. Oda and M. Tonin, {\it On the Berkovits covariant quantization   
of the GS superstring},   
Phys. Lett. B520 (2001) 398 [hep-th/0109051]\semi 
M.~Matone, L.~Mazzucato, I.~Oda, D.~Sorokin and M.~Tonin,
{\it The superembedding origin of the Berkovits pure spinor covariant  quantization of superstrings,}
Nucl.\ Phys.\ B {\bf 639}, 182 (2002)
[hep-th/0206104].
%
}  
  
\lref\kasu{
  Y.~Kazama and H.~Suzuki,
{\it New N=2 Superconformal Field Theories And Superstring Compactification,}
Nucl.\ Phys.\ B {\bf 321}, 232 (1989).
}

\lref\copeter{
G. Delius,~P. van Nieuwehuizen, and V. Rodgers, 
Int. Jour. of Mod. Phys. {\bf A5} (1990) 3943
}

\lref\Kazama{
Y.~Kazama,
{\it A Novel Topological Conformal Algebra,}
Mod.\ Phys.\ Lett.\ A {\bf 6} (1991), 1321.
}

\lref\getzler{
E.~Getzler,
{\it Manin Pairs And Topological Field Theory,}
Annals Phys.\  {\bf 237}, 161 (1995)
[hep-th/9309057].
}

\lref\karabali{
H.~J.~Schnitzer, 
{\it  A Path Integral Construction of Superconformal Field Theories from a Gauged Supersymmetric Wess-Zumino-Witten action}, Nucl. \ Phys.  B {\bf 324}, 572 (1989); 
D.~Karabali and H.~J.~Schnitzer,
{\it Brst Quantization Of The Gauged WZW Action And Coset Conformal Field Theories,}
Nucl.\ Phys.\ B {\bf 329}, 649 (1990).
}

\lref\bilal{
A.~Bilal and J.~L.~Gervais,
{\it Brst Analysis Of Super Kac-Moody And Superconformal Current Algebras,}
Phys.\ Lett.\ B {\bf 177}, 313 (1986).}

\lref\moore{
S. Ouvry, R. Stora, and P. van Baal, {\it On the Algebraic Characterization of Witten's Topological Yang-Mills theory}, 
Phys. Lett. {\bf B 220} (1989) 159; 
S.~Cordes, G.~W.~Moore and S.~Ramgoolam,
{\it Lectures on 2-d Yang-Mills theory, equivariant cohomology and topological field theories,}
Nucl.\ Phys.\ Proc.\ Suppl.\  {\bf 41}, 184 (1995)
[hep-th/9411210].}

\lref\chesterman{
M.~Chesterman,
{\it Ghost constraints and the covariant quantization of the superparticle  in ten dimensions,}
[hep-th/0212261].
}

\lref\henneaux{
M. Henneaux and C. Teitelboim, {\it Quantization of Gauge Systems}, 
Princeton, USA, Univ. Pr. (1992) }

\lref\extpure{
Y. Aisaka and Y. Kazama, {\it A new first class algebra, homological perturbation and 
extension of pure spinor formalism for superstrings}, [hep-th/0212316].
Y. Aisaka and Y. Kazama, {\it Operator Mapping between RNS and Extended Pure Spinor Formalism for Superstrings}, 
[hep-th/0305221].}

  
\Title{
\vbox{\hbox{YITP-SB-03-03 } \hbox{LPTENS 03/24} 
\hbox{DAMTP-2003-64}
}}   
{\vbox{
\centerline{The Quantum Superstring as a WZNW Model} 
\vskip .3cm
\centerline{with N=2 Superconformal Symmetry}
}}  
 
\medskip\centerline
{
P.~A.~Grassi$^{~a,}$\foot{pgrassi@insti.physics.sunysb.edu},
G.~Policastro$^{~b,}$\foot{G.Policastro@damtp.cam.ac.uk},  and  
P.~van~Nieuwenhuizen$^{~a,}$\foot{vannieu@insti.physics.sunysb.edu}
} 
\medskip   
\centerline{$^{(a)}$ 
{\it C.N. Yang Institute for Theoretical Physics,} }  
\centerline{\it State University of New York at Stony Brook,   
NY 11794-3840, USA}  
\vskip .3cm  
\centerline{$^{(b)}$ {\it DAMTP, Center for Mathematical Science, Wilberforce Road, } }  
\centerline{\it Cambridge CB3 0WA, UK}  

\medskip  
\vskip  .5cm  
\noindent  
We present a new development in our approach to the covariant quantization of superstrings 
in 10 dimensions which is based on a gauged WZNW model.  
To incorporate worldsheet diffeomorphisms we need the quartet of ghosts $(b_{zz},c^{z}, \b_{zz}, \g^{z})$ 
for topological gravity. The currents of this combined system form an $N=2$ superconformal algebra. 
The model has vanishing central charge and contains two anticommuting BRST charges, 
$Q_{S}=Q_{W} + \oint \g^{z} b_{zz} + \oint \eta_{z}$ and $Q_{V} = \oint c^{z} \Big(T^{W}_{zz} + {1\over 2} T^{top}_{zz}\Big) + \g^{z} (B^{W}_{zz} + {1\over 2} B^{top}_{zz} \Big)$, where $\eta_{z}$ is obtained by the usual fermionization of $\b_{zz}, \g^{z}$. Physical states form the cohomology of $Q_{S}+Q_{V}$, have nonnegative 
grading, and are annihilated by $b_{0}$ and $\beta_{0}$. We no longer introduce any ghosts by hand, and 
the formalism is completely Lorentz covariant. 
 
\Date{\newdate}  
 

\baselineskip16pt  

\newsec{Introduction and Summary}  

In a series of papers \grassix~we have presented a new approach to the classic problem 
of the quantization of the Green-Schwarz superstring preserving manifest super-Poincar\'e invariance in 
$D=(9,1)$. 
We began with Berkovits' formulation based on pure spinors \berkox, but we relaxed the constraints on these 
spinors by adding new ghost fields. Then we constructed a 
nilpotent BRST charge $Q$ by requiring nilpotency of the BRST transformation rules and
invariance of the free-field action (the latter requirement is equivalent to imposing holomorphicity -- or
anti-holomorphicity -- on the BRST currents: $\bar \p j_{z} =0$ is equivalent to $[Q,H]= 0$ according to the Noether 
theorem). Each time nilpotency did not hold on a given field we added 
a new ghost.  Finally, at some point we introduced by hand a ghost system $b, c_z$ which made 
the BRST charge nilpotent even though the number of fields was finite \Grassione. 
(In the past numerous approaches based on the BV formalism have led to an infinite set of ghosts \superstring.) 

The action $S$, which was BRST invariant, did not yield a vanishing central charge $c$, but introducing by hand another 
ghost pair $\omega_m, \eta^m_z$ (which was taken to be BRST inert in order not to undo the result $Q^2=0$), we also obtained $c=0$ \Grassione \Grassitwo. 
However, the resulting conformal field theory (constructed in terms of the energy-momentum tensor $T_{zz}$, 
the BRST current $j^{B}_z$, the ghost current $J_z^{\rm gh}$ and a composite antighost operator $B_{zz}$) 
could not be identified with an $N=2$ superconformal model, or with a generalization worked out 
by Kazama \Kazama~(which contains two more generators $F_{zzz}$ and $\Phi_{zzz}$ with conformal spin 
$3$ and ghost number $-2$ and $-3$, respectively). 

To obtain the correct cohomology, we required that vertex operators be BRST invariant. 
This implemented the constraints at the level of the cohomology. However, the ghost system $b, c_{z}$ 
which we had earlier introduced to obtain a nilpotent BRST operator, now turned out to be the cause that 
the cohomology was trivial. To remedy this defect, 
the concept of a grading was introduced, and vertex operators were required to have nonnegative grading \Grassitwo. 
These grading conditions were shown to be equivalent to equivariant cohomology \Grassithree.  
Homological perturbation theory \henneaux~leads to the same results, at least at the classical level (i.e., with only Poisson brackets, or with single contractions): if one removes (co)homology classes by adding new ghosts, one needs in general 
an infinite set of such ghosts \chesterman, but one may again truncate this series by introducing the $b, c_z$ 
system. The grading number turned out to be the antifield number (as defined in homological 
perturbation theory) minus the ghost number \twosix.

The same approach was shown to yield correct results for the superparticle \Grassifour, and 
even for ordinary gauge field theory \Grassifive. It was also shown how to extend this approach 
to the combined left- and right-moving sector of the superstring \Grassifour. 

In this article, we first show that if one does not short-circuit the process of 
construc\-ting a nilpotent 
BRST charge (and an invariant action) by introducing the  ghosts $b,c_z$, but instead goes on implementing  
BRST nilpotency on each field by adding more ghosts when needed, one ends up with a 
very simple system: three current multiplets $(J^g_M, J^{gh}_M, J^h_M)$ with 
$J_M=(J_m, J_\a, J^\a)$ which can be viewed as the currents of a WZNW model. Two of these multiplets, namely 
$(J^g_M, J^{gh}_M)$, contain the set of fields we found in our earlier work, whereas the third multiplet,  
$J^h_M$, is associated with the gauging of these WZNW multiplets, and contains three more currents which 
close the BRST algebra. The algebra on which this model is based is the super-Poincar\'e algebra in 10 
dimensions with a fermionic central extension. 

The currents of this model form a Kazama algebra \Kazama, the BRST current is nilpotent and the 
central charge vanishes.  
However, the corresponding BRST charge $Q_{W}$ has too much cohomology 
because vertex operators depend not only on $x^{m}, \t^{\a}, p_{z\a}$, but also 
on $x^{h,m}, \t^{h,\a}$ and $p^{h}_{z\a}$. 
Moreover, in all our work we have consistently ignored worldsheet diffeomorphisms up  till 
now. Both problems are solved by introducing a quartet $(b_{zz},c^{z}, \b_{zz}, \g^{z})$ of 
ghosts for topological gravity. The currents of this model form an $N=2$ superconformal algebra. 
The need for such a quartet was discussed in lectures by Dijkgraaf, Verlinde and Verlinde
\lref\topgrav{
J. Labastida, M. Pernici, and E. Witten, Nucl. Phys. B {\bf 310} (1989) 258; 
D. Montano and J. Sonnenschein, Nucl. Phys. B {\bf 313} (1989) 258; 
D.~Montano and J.~Sonnenschein,
{\it The Topology Of Moduli Space And Quantum Field Theory}, 
Nucl.\ Phys.\ B {\bf 324}, 348 (1989); 
R. Myers and V. Periwal, Nucl. Phys. B {\bf 333} (1990) 536; 
A. Chamseddine and D. Wyler, Phys. Lett. B {\bf 228} 75; K. Isler and 
C. Trugenberger, {\it A Gauge Theory Of Two-Dimensional Quantum Gravity,}
Phys.\ Rev.\ Lett.\  {\bf 63}, 834 (1989); E.~Witten, {\it On The Structure of the Topological Phase 
of Two-Dimensional Gravity,}
Nucl.\ Phys.\ B {\bf 340}, 281 (1990).} 
\lref\dvv{ 
R.~Dijkgraaf, H.~Verlinde and E.~Verlinde,
{\it Notes On Topological String Theory And 2-D Quantum Gravity,}
PUPT-1217
{\it Based on lectures given at Spring School on Strings and Quantum Gravity, Trieste, Italy, Apr 24 - May 2, 1990 and at Cargese Workshop on
Random Surfaces, Quantum Gravity and Strings, Cargese, France, May 28 - Jun 1, 1990}
}
\dvv (see also \topgrav); 
it provides the parametrization of the moduli of the Riemann surfaces. 
Combining our WZNW model with the topological gravitational quartet, 
the properly modified currents of this combined system form an $N=2$ superconformal algebra 
\figueB. 
One of these currents is the BRST current $j_{S}^{B} = j^{B}_{W} + j^{B}_{top}$, where 
$j^{B}_{W}$ is the BRST current of the WZNW model while $j^{B}_{top}=\g^{z} b_{zz}$ is the BRST 
current of the topological gravity. In addition, as for any topological model, there is a second BRST 
charge which anticommutes with $Q_{S}$, given by \dvv  
\eqn\bbrrsstt{
Q_{V} = \oint c^{z} \Big(T^{W}_{zz} + {1\over 2} T^{top}_{zz}\Big) + \g^{z} (B^{W}_{zz} + 
{1\over 2} B^{top}_{zz} \Big)\,.
}

The main  point of this article is the definition and construction of physical states. 
Physical states correspond to vertex operators which are polynomials in the fields and derivatives thereof, 
lie in the cohomology of ${Q}_{S} + \oint \eta_{z} + Q_{V}$, have nonnegative grading, 
and are annihilated by the zero modes $b_{0} = \oint z b_{zz}$ and $\b_{0} =\oint z \beta_{zz}$. The field $\eta_{z}$ is obtained by 
fermionizing the commuting ghosts $\b_{zz}$ and $\g^{z}$, and the 
operator $\oint \eta_{z}$ is added to the BRST charge 
$Q_{S}+Q_{V}$ for the same reasons for which it is added in the RNS 
formalism \bvw\  working in the large Hilbert space.
Since $\Big\{b_{0}, Q_{S} + Q_{V} +\oint \eta_{z}\} = \oint z (T^{W}_{zz} + T^{gh}_{zz})$ and 
$\Big[ \beta_{0}, Q_{S} +Q_{V} + \oint \eta_{z} \Big] =b_{0} + \oint z B_{zz}$, 
the requirement that $b_{0}$ and $\beta_{0}$ annihilate vertex operators puts them on-shell, 
and (as we shall show) eliminates the doubling in the WZNW model mentioned above. 

Thus we have found a covariant formulation of the quantum superstring with the following 
properties

\noindent
{\it 1 )} it is based on a WZNW model, 

\noindent
{\it 2 )} it is conformally invariant (it has vanishing central charge),

\noindent
{\it 3 )} the currents form an $N=2$ conformal superalgebra , 

\noindent
{\it 4 )} it yields the correct cohomology (checked for the open string 
in the sector with ghost number one and conformal spin zero, to be published elsewhere), 

\noindent
{\it 5 )} no ghosts are any longer introduced by hand. 

Having shown that our previous work is based on a WZNW model suggests that 
covariant quantum computations in superstring theory may be easier than thought. 
In particular, the precise form of the measure, which is crucial for tree and loop level 
computations, may be easier to determine for this formulation with WZNW currents. 
Perhaps this very general and minimal model can finally shed light on the mysterious classical 
$\kappa$-symmetry. 

Before concluding this introduction, we would like to mention related work. 
The Padua group has given a derivation of the pure spinor formalism using
the complexified superembedding of the twistorial version of GS superstrings \tonin. 
They showed that the pure spinor formalism originates from the superembedding approach to superbranes, 
namely it arises as  a result of a conventional BRST gauge-fixing of a complexified and twisted $N=2$ worldsheet supersymmetric superembedding of the Green-Schwarz superstring. 
In the light of our developments, it might be useful to compare the two 
formalisms to construct the underlying classical gauge invariant action. 

An approach to pure spinors which differs from that of Berkovits was begun by  
Aisaka and Kazama \extpure. 
They factorized the pure spinor constraints into a reducible and irreducible 
part preserving the subgroup $U(5)$ of the Lorentz group. This factorization leads to 
a new first class algebra of constraints which yields the BRST charge by the usual construction. 
The set of new ghost fields needed for implementing the constraints has vanishing conformal charge. It would be interesting  to compare their formalism with the results of the present work where we use the gauged WZNW model to obtain first class constraints. 

The paper is organized as follows: in section 2, we discuss the underlying WZNW 
structure, we derive the Maurer-Cartan forms and we introduce the $h$-sector of 
currents. In section 3, we construct the BRST invariant action and point out 
its relation to the classical Green-Schwarz action and the free field action on which 
our earlier work is based. In section 4, the 
underlying conformal field theory is analyzed. In section 5 the quartet for topological gravity is introduced, 
and its current algebra derived. Finally, in section 6 the definition
of physical states is given. The relation to our previous approach is given in section 7.
In the conclusions, we discuss open issues and possible future applications. 


\newsec{The underlying WZNW structure}

We start from Berkovits' BRST-like charge $Q_{B}= \oint dz \, j_z^B(z)$ with $j^B_z(z) = 
i\, \l^\a d_{z \a}$ and $\l^{\a}$ ($\a =1,\dots,16$) a real commuting spinor-ghost, 
but we do not impose the constraints $\l \g^m \l =0$. We follow the notation and the definitions of \Grassisix. 
The operator $d_{z\a} = p_{z\a} + i \p_z x^m  (\g_{m} \t)_{\a} + 
{1\over 2} (\g^m \t)_\a \t \g_m \p_z \t$ (we restrict ourselves to the left-moving sector) yields 
 BRST transformations on $x^m$ and $d_{z\a}$ (using $x^{m}(z) x^{n}(w) \sim - \eta^{mn} \ln(z-w)$ and 
 $p_{z \a}(z) \t^{\b}(w) \sim \delta_{\a}^{\b} /(z-w)$) 
 which are not nilpotent, but become nilpotent if one adds 
 ghosts $\xi^m$ and $\chi_\a$ (and antighosts $\beta_{z m}$ and $\kappa^\a_{z}$; the antighost for $\l^\a$ is 
 $w_{z \a}$). The resulting antihermitian BRST charge reads \Grassione
 \eqn\BRSTA{
Q = \oint dz \Big( i \l^\a d_{z \a} - \xi^m \Pi_{z m} - \chi_\a \p_z \t^\a 
 - 2 \xi^m (\k_{z} \g_m \l) - i \beta_{z m} \l \g^m \l
 \Big)\,,}
 with $\Pi^{m}_{z} = \p_{z} x^{m} - i \t \g^{m} \p_{z} \t$. 

The first three terms in $Q$ can be written as $ - J^{g}_{M} c^{M} $
with $c^{M} = ( \xi^{m},  \l^{\a}, \chi_{\a})$ and $J^{g}_{M} =(\Pi_{z m}, - i d_{z \a}, \p_{z} \t^{\a})$. 
We view the $J^{g}_{M}$ which appear in $Q$ as operators whose classical counterparts 
are first class constraints which determine the structure constants. 
From the OPE's 
$$
(- i d)_{z\a}(z) (-i d)_{w\b}(w) \sim   - 2 i {\g^m_{\a\b} \Pi_{w m}(w) \over z-w}
\,, 
~~~~~
(-i d)_{z\a}(z) \Pi^{z}_{m}(w) \sim - 2  {\g_{m,\a\b} \p_{w}\t^{\b} (w)\over z-w}\,, 
$$ 
\eqn\gcurr{ 
\Pi_{z m}(z) \Pi_{w n}(w) \sim   - {1 \over (z-w)^2} \eta_{mn}\,,~~~~~~~ 
(-i d)_{z\a}(z) \p_w \t^{\b}(w) \sim  - { i  \over (z-w)^2} \delta^{~\b}_{\a}\,, 
}
we can extract an affine Lie algebra
\eqn\aff{
J^{g}_{M}(z) J^{g}_{N}(w) \sim {J^{g}_{P} f^{P}_{~MN} \over (z-w)} - {{\cal H}_{MN} \over (z-w)^{2} }\,. 
}
Conversely, requiring closure of the affine Lie algebra fixes the trilinear term in $d_{z\a}$.

Introducing abstract generators $T_{M} = (P_{m}, Q_{\a}, K^{\a})$ satisfying $[T_{M} , T_{N} \} = 
T_{P} f^{P}_{~MN}$, we find only two nonvanishing structure constants
\eqn\stru{
\{Q_{\a}, Q_{\b} \} = - 2 i \g^m_{{\a\b}} P_{m}\,, ~~~~~~
[Q_{\a}, P_{m} ] =  - 2  \g_{m,{\a\b}} K^{\b}\,. 
}
Introducing the antighosts $b_{M} = (\beta_{z m},   w_{z \a},  \, \kappa_{z}^{\a})$ 
satisfying $c^{M}(z) b_{N} (w) \sim \delta^{M}_{~N} {1\over z-w}$, the BRST charge in \BRSTA~can be written as 
\eqn\struA{
Q = \oint dz \Big( -J^{g}_{M} c^{M}  - {1\over 2} b_{M} f^{M}_{~NP} c^{P} c^{N} (-)^{N} \Big)
=
- \oint dz \Big( J^{g}_{M} + {1\over 2} J^{gh}_{M} \Big) c^{M}\,,
}
where $(-)^{N} = +1$ for $P_{m}$ and $(-)^{N} = -1$ for $Q_{\a}, K^{\a}$.  The ghost 
currents 
\eqn\ghocurre{
J^{gh}_M=b_{M} f^{M}_{~NP} c^{P} (-)^{N} = 
\Big( 2 \k_z \g_m \l, 2 \xi^m (\g_m \k_z)_\a + 2 i \beta_{z m} (\g^m \l)_\a, 0 \Big)\,,
}
satisfy \aff~without double poles. 

At this point, we make contact with work by Green and Siegel of a decade ago.
In \csm~it was shown that the Green-Schwarz action can be reformulated as a WZNW model based 
on a super Lie algebra with abstract generators $P_{m}, Q_{\a}, K^\a$,  which correspond to the left-invariant one-forms 
$ \Pi^m_z$, $\p_{z} \t^\a$, and $d_{z\a}$ appearing in our work. 
This super Lie algebra is an extension of the usual $D=(9,1)$ super-Poincar\'e algebra with $Q_\a$ and $P_{m}$ 
to a super Lie algebra where $K^\a$ is a central charge. It is nilpotent, but has a non-degenerate invariant metric \grr.  
The three current multiplets $(J^g_M, J^{gh}_M, J^h_M)$ form representations of
this super Lie algebra. In the next paragraph we give some details. 

 A coset approach with unitary $g = e^{ P_{m} x^{m} } e^{  Q_{\a}\t^{\a}} e^{ K^{\a} \phi_{\a}}$, 
 containing antihermitian $P_{m}$, hermitian $Q_{\a}$ and $K^{\a}$, 
 satisfying \stru, 
 and real $x^{m}, \t^{\a}$ and $\phi_{\a}$, leads to the usual left-invariant one-forms, 
 Lie derivatives and covariant derivatives. 
 In particular the left-invariant one-form $g^{-1} d g$ corresponding to $Q_{\a}$ is equal to 
 $\p_{z} \t^{\a}$, and the one-form corresponding to $P_{m}$ is given by $\Pi^{z,m} = 
 \p_{z} x^{m} - i \t \g^{m} \p_{z}\t$.  The one-form $g^{-1} d g$ corresponding to
 $K^{\a}$ yields the current $ - i d_{z \a} = \p_{z} \phi_{\a} +  
 2 \, \p_{z} x^{m} (\g_{m} \t)_{\a} -  {  2 i \over 3} (\g_{m} \t)_{\a} (\t \g^{m} \p_{z} \t)$. 
 Defining $p_{z \a} =  i \p_{z}\phi_{\a} +i \p_{z} x^{m } (\g_{m}\t)_{\a} + {1\over 6} (\g_{m} \t)_{\a} (\t \g^{m} \p_{z} \t)$ 
one obtains the operator $d_{z \a}$ appearing in (2.1), $d_{z\a} = p_{z\a} + i \p_z x^m  (\g_{m} \t)_{\a} + 
{1\over 2} (\g^m \t)_\a \t \g_m \p_z \t$ where $p_{z\a}$, $\t^{\a}$ and $x^{m}$ are free fields (see below). 
  The Lie derivative $K^{\a} = \p / \p \phi_{\a}$ can be represented as  
 $K^{\a} = - i  \oint dz \p_{z} \t^{\a}$ when $\t^{\b}(z) \phi_{\a}(w) \sim i \delta_{\a}^{~\b} ln (z-w)$, but it vanishes (being 
 an integral of a total derivative) when it acts in the space which contains only $\p_{z} \phi_{\a}$.  
 This is due to the fact that $K^{\a}$ generates constant 
 shifts of the coordinate $\phi_{\a}$, but  only $\p_{z} \phi_{\a}$ (identified with the conjugate  momentum of $\t^{\a}$) appears in the theory. 

The generator for rigid spacetime supersymmetry $\oint dz\, q_{z \a}$ can be determined by requiring that it leaves 
$\Pi^{m}_{z}, \p_{z} \t^{\a}$, and $d_{z\a}$ invariant. It is given by $q_{z \a} = p_{z\a} - i \p_{z} x^{m } (\g_{m}\t)_{\a} 
- {1\over 6} (\g_{m} \t)_{\a} (\t \g^{m} \p_{z} \t)$ \Grassione, which takes on a very simple form in terms of $x,\t$ and $\phi$, namely 
$q_{z\a} = i \p_{z} \phi_{\a}$. Since we only use composite operators which are susy invariant, 
$N=(1,1)$ spacetime susy is manifestly maintained at all stages.

The currents $J^{g}_{M}(z)$ in \struA\ are related to the left-invariant one-forms $g^{-1} d g = T_{M} J^{g, M}$ 
given by $J^{g,M} = (\Pi^{m}_{z}, \p_{z}\t^{\a}, - i d_{z\a})$. 
They satisfy $J^{g}_{M} = H_{MN} J^{g,N}$ where $H_{mn} = \eta_{mn}$, $H_{\a}^{~\b} = 
H^{\b}_{~\a} = \delta_{\a}^{\b}$. 
The matrix $H_{MN}$ is not an invariant metric. The easiest way to see this is to take its inverse $H^{MN}$ and 
to construct the bilinear expression $T_{N} T_{M} H^{MN} = P_{m} P^{m}$; since this operator does not 
commute with all generators, the matrix $H^{MN}$ is not an invariant matrix. One can understand this 
by noting that the explicit expressions for the left-invariant one-forms depend on the basis for the group. Taking 
for example $g = e^{Q_\a \t^{\a}} e^{P_{m} x^{m}} e^{K^{\a} \phi_{\a}} $ one finds\foot{Rewriting this group element 
as $g =  e^{P_{m} x^{m}} e^{Q_\a \t^{\a}} e^{K^{\a} (\phi_{\a} + 2 x ^{m} (\g_{m}\t)_{\a})} $, the result in (2.7) follows.} 
\eqn\cca{
g^{-1} d g = 
P_{m} \Pi^{m} + Q_{\a} d \t^{\a} + K^{\a} 
\Big[d \phi_{\a} - 2 \, x^{m} (\g_{m} d\t) - {2 i \over 3} (\g^{m} \t)_{\a} (\t \g_{m} d \t)\Big]\,.
} 
This expression differs from $T_{M} J^{g,M}$, note the bare $x$. 
For given $J^{g}_{M}$ and different parametrization of $g$ also  $J^{g,M}$ will be different, and in this sense 
$H_{MN}$ is basis-dependent. Taking different bases is like a choosing a gauge: the physical results 
(cohomology) should be basis-independent. 

In order to construct the WZNW action we do need an invariant metric.
One can find an invariant metric ${\cal H}^{MN}$ by constructing the Casimir 
operator\foot{One may check that the sign $(-)^{N}$ in the Casimir operator is needed by working out 
the case of $Osp(1|2)$. The generators are the bosonic generators $(J_{3}, J_{+}, J_{-})$ of 
$Sp(2) = SU(1,1)$, and the two fermionic generators $(Q_{+}, Q_{-})$ . They satisfy 
$[J_{3}, J_{+}] = J_{+}$, $[J_{3}, Q_{+}] = {1\over 2} Q_{+}$, $[J_{+}, J_{-}]= 2 J_{3}$ and 
$[J_{\pm}, Q_{\mp}] = - Q_{\pm}$. The anticommutators are $\{Q_{+}, Q_{-} \} = {1\over 2} J_{3}$ and 
$\{Q_{\pm}, Q_{\pm} \} = \pm {1 \over 2} J_{\pm}$. The Casimir operator is 
$C_{2} = J_{3}^{2} + {1\over 2} (J_{+} J_{-} + J_{-} J_{+} ) - (Q_{+} Q_{-} - Q_{-} Q_{+})$. Assuming that 
$C_{2}$ is given by the formula in the text with the factor $(-)^{N}$ one finds on the basis 
$(J_{3}, J_{+}, J_{-}, Q_{+}, Q_{-})$ that ${\cal H}^{MN}$ is the block-diagonal matrix 
$(1, {1\over 2} \tau_{1}, -i \tau_{2})$, and thus ${\cal H}_{MN}$ is the block-diagonal matrix 
$(1, 2 \tau_{1},  i \tau_{2})$. This agrees with the Killing metric ${\cal H}_{MN} 
=c\,  f^{P}_{~MQ} f^{Q}_{~NP} (-)^{P}$ with $c=2/3$. 
(The factor $(-)^{P}$ in this expression is needed because the contraction of the indices $P$ does not follow our northeast-southwest convention). The only simple superalgebras with nondegenerate 
Killing metric are $SU(m|n)$ for $m\neq n$, $Osp(m|n)$ except $Osp(2m|2m+2)$, and $F(4)$ and $G(3)$ 
\lref\FrappatPB{
L.~Frappat, P.~Sorba and A.~Sciarrino,
{\it Dictionary on Lie superalgebras,} hep-th/9607161.
}
\FrappatPB. In
our case we are of course dealing with a nilpotent algebra (all triple-(anti)commutators vanish), 
but we have explicitly exhibited an invariant metric.  
} 
$C_{2} = T_{N} T_{M} {\cal H}^{MN}(-)^{N}$. It is in our case given by $P^{m} P_{m} + 2 i Q_{\a} K^{\a}$.
One can also use the metric ${\cal H}_{MN}$ which is given by the central charges in \aff. This is also an invariant 
metric and it is the inverse of the metric in the Casimir operator as we now show. Defining an inner product 
for generators $(T_{M}, T_{N}) = {\cal H}_{MN}$, invariance of this inner product under adjoint transformations 
leads to the relation $-([T_{M}, T_{P}], T_{N}) + (T_{M}, [T_{P}, T_{N}]) =0 $. In terms of ${\cal H}_{MN}$ this 
reads $- {\cal H}_{RN} f^{R}_{~MP} + {\cal H}_{MR} f^{R}_{~PN} =0 $. Raising indices with ${\cal H}^{MN}$ 
yields $- f^{A}_{~PM} {\cal H}^{MB} + f^{B}_{~PN} {\cal H}^{NA} =0$, and this equation is also the equation 
one gets if one requires that $C_{2}$ commutes with $T_{P}$. The final result reads
\eqn\metric{
{\cal H}^{MN}  =  
\left( 
\eqalign{ 
&\eta^{mn} ~~~~~ 0 ~~~~~~~~~ 0            \cr 
& 0             ~~~~~~~~~ 0 ~~~~~~~~~  i \d^\a_{~\b}  \cr 
&0 ~~~~~~~ -i \d_\a^{~\b} ~~~~~~ 0  
} \right)\,,
~~~~~~~~~
{\cal H}_{MN} =   
\left( 
\eqalign{ 
& \eta_{mn} ~~~~~ 0 ~~~~~~~~~ 0            \cr 
& 0             ~~~~~~~~~ 0 ~~~~~~~~~  i \d_\a^{~\b}  \cr 
&0 ~~~~~~ - i \d^\a_{~\b} ~~~~~~ 0  
} \right)\,.
}
For semisimple super Lie algebras one may use the supertrace to construct an invariant metric: 
$str( J c) = str(T_{M} T_{N}) J^{N} c^{M}$ where 
the supertrace $str(T_{M} T_{N}) \equiv {\cal H}_{MN}$ is nonvanishing. 
We have not found an anti-de Sitter extension for the set $T_{M}$, so we could not have used a 
Wigner-In\"onu contraction to obtain ${\cal H}_{MN}$.

For our considerations it is useful to keep track of reality properties. 
Note that $J^{g}_{M}$, $J^{g,M}, c^{M}$, and $J^{g}_{M} c^{M}$ are hermitian,  $T_{M} J^{g, M}$ 
is antihermitian, but $c = T_{M} c^{M}$ and $b_{M}$ have no definite reality properties. The ghost 
currents $J^{gh}_{M}$ have the same reality properties as the gauge currents $J^{g}_{M}$. 
  
Classically (i.e., taking only single contractions into account) $Q$ is nilpotent. 
However, it fails to be nilpotent when acting on the antighosts $b_{M}$. 
This is due to the double poles in the current algebra 
(due to derivatives of simple contractions: there are no double contractions) generated 
by $J^g_M$,\foot{More specifically, the BRST transformation of $\kappa^{\a}_{z}$ is 
proportional to $J^{g,\a}_{z}$ but the BRST transformation of $J^{g ,\a}_{z}$ is due to the double pole in 
$J^{g}_{z \a}(z) J^{g,\b}_{z}(w)$. Similarly for $\b^{m}_{z}$ and $w_{z \a}$.} whereas the current algebra generated 
by the ghost currents $J^{gh}_M$ 
does not have double poles\foot{For a generic WZNW model the ghost currents $J^{gh}_M$ produce 
double poles due to double contractions and these are needed to cancel the double poles 
of the other currents $J^g_M  + J^h_M$. For $J^{gh}_{M}$ 
only the OPE $J^{gh}_{z \a}(z) J^{gh}_{z \b}(w) \sim - 2 i \g^{m}_{\a\b} J^{gh}_{z m} / (z-w) $ is nonvanishing. 
}. 
Following the usual treatment of gauged WZNW models and the corresponding susy coset models \karabali, we introduce new hermitian 
currents\foot{Recall that 
integrating out the gauge fields $A_z$ and $\bar A_{\z}$ for the diagonal maximal subgroup, one obtains a Jacobian 
which can be exponentiated to yield another WZNW model for this subgroup \karabali.} 
\eqn\BRSTB{
J^h_M = (J^h_{z m}, J^h_{z \a}, J^{h, \a}_{z})
}
which correspond to $J^{g}_{M} = (\Pi_{z m },  - i d_{z\a}, \p_{z}\t^{\a})$. 
We determine their transformation rules in the following way: we add these new currents to the 
BRST transformation laws of the antighost fields and then we require nilpotency on the antighosts. From 
$$
[ Q, \k_{z}^\a] = -\p_z \t^\a - J^{h, \a}_{z} \,,
$$
$$
\{ Q, \b_{z m} \} = -\Pi_{z m} - 2 \k_z \g_{m}\l -  J^{h}_{ z m} \,,
$$
\eqn\BRSTC{
[Q, w_{z\a} ] = i \, d_{z\a} - 2 i \b_{z m}\, (\g^m\l)_\a - 2 \, \xi^m(\g_m \k_z)_\a - J^h_{z \a}\,,
}
we obtain
$$
\{ Q, J^{h, \a}_{z} \} = - i \p_z \l^\a \,,~~~~~~
[ Q, J^{h}_{z m} ] = - \eta_{mn} \p_z \xi^n + 2\, J^{h, \a}_z \g_{m, \a\b} \l^{\b} \,,
$$
\eqn\cicA{
\{Q, J^h_{z\a} \} =  i \p_{z} \chi_\a  + 2 i  J^h_{z m}\, (\g^m\l)_\a  - 2  \, \xi^m(\g_m J^h_z)_\a \,.
}
It turns out that without further ghost fields and generators all BRST transformations are also nilpotent on the 
$h$-currents. Hence, the BRST transformations are now nilpotent on all fields, and  
the iterative construction which we followed to obtain a nilpotent BRST charge, terminates at this early point.
The result is $Q = \oint dz j^{B}_{z}$ with
 $$
j^{B}_{z} =  - \l^\a \Big( -i d_{z \a} + J^h_{z\a}\Big) - \, \xi^m \Big(\Pi_{z m} + J^h_{z m}\Big)
 - \, \chi_\a \Big(\p_z \t^\a + J^{h,\a}_{z}\Big) + 
$$
\eqn\cicB{
 - 2 \xi^m (\k \g_m \l) - i \beta_{z m} \l \g^m \l \,.
}
From \cicA~and requiring nilpotency of \cicB~we deduce that 
the only nontrivial OPE's of the $h$-currents are given by
$$
J^h_{z\a}(z) J^h_{w\b}(w) \sim  - 2 i 
{\g^m_{\a\b} J^h_{w m}(w) \over z-w}
\,, 
~~~~~
J^h_{z\a}(z) J^h_{w m}(w) \sim - 2 
{\g_{m,\a\b} J^{h\b}_{w}(w)\over z-w 
}\,, 
$$ 
\eqn\hcurr{ 
 J^h_{z m}(z) J^h_{w n}(w) \sim  {1 \over (z-w)^2} \eta_{mn}\,,~~~~~~~ 
J^h_{z\a}(z) J^{h\b}_w(w) \sim   {i \over (z-w)^2} \delta^{~\b}_{\a}\,. 
} 
Note that the sign of the double poles in the OPE's for $J^{h}_{M}$ is opposite to the sign 
of the double poles for the currents $J^g_M$. This will play a role below. 


\newsec{The action}

The BRST charge $Q$ obtained above corresponds for the left-moving sector to the BRST charge for a 
WZNW  model on $G \times G / G$ where the maximal diagonal subgroup G is gauged. 
$G$ is the superalgebra in $D=3,4,6,10$ dimensions generated by $(P_m, Q_\a, K^\a)$. 
 
 The action for this WZNW model is given by
$$
 S_{g} = \int d^{2}x {1\over 2} \eta^{\mu\nu} J^{g}_{\mu M} J^{g}_{\nu N} {\cal H}^{NM} + 
 k \int d^{3}x \e^{\mu\nu\rho} J^{g}_{\mu M} J^{g}_{\nu N}  J^{g}_{\rho R} {f}^{RNM} =
$$ 
$$
= \int d^2x \left( {1\over 2} \eta_{mn} J^{g,m}_\mu J^{g,n}_\nu - i \, J^{g}_{\a \mu} J^{g,\a}_\nu  
\right) \eta^{\mu\nu}  
 $$
 \eqn\WZW{
 + 2\, i \int d^3x \e^{\mu \nu \rho} 
 \g_{m\a\b} \left( J^{g,m}_\mu J^{g,\a}_\nu J^{g,\b}_\rho 
 \right)  \,,
 }
where the indices of the structure constants have been raised with the invariant metric and the 
constant $k$ has been chosen such that $\t$ will be left-moving. 
 Since the Maurer-Cartan equations $d J^{g,M} =  - {1\over 2} f^M_{~~NP} J^{g,P} J^{g,N}$ imply that 
 $2\, \g_{m \a\b} J^{g,m}_{[\mu} J^{g,\a}_{\nu]}  =  - \p_{[\m} J^{g}_{\n]\b}$, the second term in $S_{g}$ can be written as a two-dimensional integral, and the two terms with $\eta^{\m\n}$ and 
 $\e^{\m\n}$ combine into the chiral combination $ i J^{g}_{z \a} J^{g, \a}_{\z}$. We can, of course, always add a term, 
 $- {1\over 2}\eta_{mn} J^{g,m}_\m J^{g,n}_\n \e^{\m\n}$ to the action as it vanishes, and obtain then also a chiral 
 expression for the terms with $J^{g,m}_{z}$. The result 
 reads 
 \eqn\WZWa{
 {\cal L}_{g} = {1\over 2} \Pi^m_z \Pi^n_{\bar z} \eta_{mn} + d^{(\phi)}_{z\a} \bar\p \t^\a 
 \,,}
where $d^{(\phi)}_{z\a} = i J^{g}_{z\a}$ 
still depends on $\phi_{\a}$, see section 2. 

We now replace $J^{g}_{z \a}$ by $ -i d_{z\a}$ which amounts to replacing $ \p_{z} \phi_{\a}$ by 
$ -i p_{z\a} + \dots$, as explained in section 2.
 Substituting the explicit expressions for $d_{z \a}$ and $\Pi^m_z$, the action becomes the free-field action 
 from which we started in our previous work
 \eqn\WZWa{
 {\cal L} = {1\over 2} \p_{z} x^m \bar \p_{\z} x^n \eta_{mn} + p_{z\a} \bar\p_{\z} \t^\a 
 \,,
 }
with $\p_{z} = \p_{\sigma} - i \p_{\tau}$ and $\bar\p_{\z} = - \p_{\sigma} - i \p_{\tau}$. 
We therefore discover at this point that our previous work \grassix~was based on a WZNW model. 
The original WZNW model in terms of $\phi, \theta$ and $x$ is a complicated interacting theory, but 
by introducing the variable $p_{z\a}$ (also a complicated expression in terms of $\phi,\t$ and $x$), 
one obtains a free-field action. Thus $p,\t$ and $x$ form a free-field realization of the affine Lie algebra. 

The replacement of $\p_{z}\phi_{\a}$ by $- i p_{z \a} + \dots$ can be justified as follows. The fields in the 
BRST operator are on-shell and on-shell $\bar\p_{\z} \phi_{\a} = 0$. In that case one can solve $\phi_{\a}$ in terms 
of $p_{z\a}$ and replacing $\phi_{\a}$ by $p_{z\a}$ amounts to a change of basis. 
In the WZNW action \WZW, on the other hand, $\bar\p_{\z} \phi_{\a}$ is nonvanishing, but $\phi_{\a}$ only  
appears in the combination $\p_{\mu} \phi_{\a} \p^{\mu} \t^{\a}$. This expression may again be replaced by 
$p_{\mu \a} \p^{\mu} \t^{\a}$ since $p_{\mu \a}$ can be decomposed into a gradient $\p_{\mu} \phi_{\a}$ and 
a curl $\epsilon_{\mu\nu} \p^{\nu} \phi_{\a}$, and the latter is pure gauge.   

Our approach also gives a geometrical interpretation of Berkovits' approach. Gauging the 
generator $Q_\a$ is classically equivalent to setting the current $J^{g}_{\a} = - i d_\a$ 
in \WZW~equal to zero, and this 
yields the classical Green-Schwarz action. A more group theoretical way to set $J^{g}_{\a}=0$ 
involves a group contraction.  
The Lie algebra generated by $P_{m}, Q_{\a}, K^{\a}$ has an outside automorphism $G$
\eqn\grading{
[G, Q_{\a}] = {1\over 2} Q_{\a}\,, ~~~~
[G, P_{m}] = P_{m}\,, ~~~~
[G, K^{\a}] = {3\over 2} K^{\a} \,.
}
It leads to the grading of the ghosts $\l^{\a}$, $\xi^{m}$ and $\chi_{\a}$ which we used in our 
earlier articles to define the cohomology \Grassitwo. 
In addition, one can introduce a contraction parameter $R$ as follows 
\eqn\defalg{
\{Q_{\a}, Q_{\b}\} =  - 2 i \g^{m}_{\a\b} P_{m}\,, ~~~~~~~
[ Q_{\a}, P_{m}] = -2 \, R\, \g^{m}_{\a\b} K^{\b} \,. 
}
The currents $J^{g,M}$ become now $R$-dependent, but evaluating the Wess-Zumino term ${\cal H}_{RP} f^{P}_{~MN} J^{g,N} J^{g,M} J^{g,R}$ one finds that it is R-independent. In the kinetic term ${\cal H}_{MN} J^{g,N}_{\mu} J^{g,M}_{\nu} \eta^{\m\n}$, the one-form associated with $K^{\a}$ becomes $R$-dependent 
\eqn\newd{
d_{z\a} = i \p_{z} \phi_{\a} +  2 i R \, \p_{z} x^{m} (\g_{m} \t)_{\a} + {  2 R \over 3} (\g_{m} \t)_{\a} (\t \g^{m} \p_{z} \t)\,,
}
while ${\cal H}_{MN}$ acquires a factor $1/ R$ in the fermionic sector. If one defines $p_{z \a}$ 
such that $d_{z\a}$ becomes $R$-independent (for example by choosing for $p_{z\a}$ the expression given in section 1), 
one can take the limit $R \rightarrow \infty$ and obtains the classical Green-Schwarz action. 

At the quantum level the constraint $d_\a =0$ is implemented 
by Berkovits' BRST charge $Q_B = \oint i\, \l^\a d_\a$. The condition $Q_B |\psi \rangle =0$ would 
be  the natural condition 
for gauging $J_\a$, but since $Q_B$ is not nilpotent, one must impose the pure spinor constraint $\l \g^m \l =0$ in his approach. 
In conventional gauged WZNW models one can only gauge a subalgebra. 
In the present case one can only gauge $K^\a, P_{m}$ or $\{K^\a, P_{m}\}$ in each sector of $G \times G$ 
since only they generate a proper Lorentz-invariant subalgebra, 
or the whole of the diagonal subgroup $G$ in $G\times G$.  
Taking the latter case the gauging of $G$ leads to the multiplet of currents $J^{h}_{M}$, and as action 
for these currents we take  
\eqn\haction{
S_h = - \int d^2z \Big( {1\over 2} J^h_{z m} J^{h m}_\z + i\, J^h_{z\a}  J^{h\a}_{\z}  
\Big)\,.
} 
The minus sign in front of this action 
amounts to changing the level $k=1$ into $k=-1$. The propagators of $x^{h}, \t^{h}$ and $p^{h}_{z}$ have an 
extra minus sign, and if the $h$-currents in terms of $h$-coordinates differ from 
the corresponding $g$-currents by an extra overall minus sign, then one obtains (2.13) with the same 
structure constants as for the $g$-currents, but with an extra minus sign for the double poles. 
Without the currents $J^{h}_{M}$ nilpotency of $Q$ requires further terms depending on the ghosts 
$b$ and $c_{z}$, but with $J^{h}_{M}$ the double poles due to $J^{g}_{M}$ are cancelled, and no 
$b,c_{z}$ terms are needed for nilpotency. 


\newsec{Conformal Field Theory}

Having obtained a WZNW formulation, we can study its properties as a conformal field 
theory. One can construct the energy-momentum tensor $T_{zz}$ 
\eqn\emT{
T_{zz} = - {1\over 2} \p_{z} x^{m} \p_{z} x_{m} - p_{z \a} \p_{z} \t^{\a} - \b_{z m} \p_{z} \xi^{m} - \k^{\a}_{z} \p_{z} \chi_{\a} - w_{\a} \p_{z} \l^{\a}}
$$
+{1\over 2} J^{h}_{z m} J^{h, m}_{z} + i \, J^{h}_{z \b} J^{h,\b}_{z}  \,.
$$
The first two terms can be rewritten as $ - {1\over 2} \Pi_{m z} \Pi^{m}_{z} - d_{\a} \p_{z} \t^{\a}$. (Since the Killing-Cartan 
metric vanishes, the prefactor in the  Sugawara construction of the energy-momentum tensor in the 
$h$-sector equals unity.)
Since the action is a free action, it is easy to check that the conformal charge is zero. Explicitly: 
the $c = 10 -32 -20 + 32 + 32 = 22$ 
of the sector with $J^{g}_{M}$ and $J^{gh}_{M}$ is cancelled by the $c= 10 -32$ of the sector 
with $J^{h}_{M}$. The ghost current is given by
\eqn\ghocuurrentrr{
j^{gh}_{z} = - \b_{z m} \xi^{m} - \kappa_{z}^{\a} \chi_{\a} -  w_{z \a} \l^{\a} \,.
}
Since 
the anomaly in the OPE $j^{gh}_z(z) j^{gh}_w(w) = c_j/(z-w)^2$ of the ghost current with itself 
is not zero but given by $c_j = - 22$, while $T_{zz}(z) j^{gh}_{w}(w) = 22/ (z-w)^{3} + j^{gh}_{z}(z)/(z-w)^{2}$, 
this superconformal algebra seems to be twisted. 

In addition to the BRST current $j^{B}_{z}$ given in \cicB\ and satisfying 
\eqn\nilJ{
j^{B}_{z}(z) j^{B}_{w}(w) \sim 0\,, 
 }
there is another fermionic operator $B_{zz}$ dual to the BRST current, 
obtained by interchanging ghosts and antighosts \bilal\ 
and taking the difference of the $g$-currents and $h$-currents \figue 
\eqn\bigbb{
B_{zz} = -{i\over 2}  \k^\a_z \Big( -i d_{z \a} - J^h_{z\a}\Big) + {1\over 2} \b^m_z \Big(\Pi_{z m} - J^h_{z m}\Big)
 + {i\over 2} w_{z\a} \Big(\p_z \t^\a - J^{h,\a}_{z}\Big) \,.
 }
It is an antihermitian spin 2 operator, and satisfies 
\eqn\jb{
j^{B}_{z}(z) B_{ww}(w) \sim  
{ -22 \over (z-w)^{3}} +
{j^{gh}_{w}(w) \over (z-w)^{2}} + {T_{ww}(w) \over (z-w)}\,, ~~~~~ T_{zz}(z) = \{ Q, B_{zz}(z) \}\,, }
as well as 
\eqn\bb{
B_{zz}(z) B_{ww}(w) \sim {F_{www} \over (z-w)} \,,}
where 
\eqn\bbF{
 F_{zzz}(z) = - i \b^{m}_{z} \k_{z} \g_{m} (\p_{z} \t^{\a} + J^{h\a}_{z}) + 
{i \over 2} (\k_{z} \g_{m} \k_{z}) (\Pi^{m}_{z} + J^{h m}_{z})   \,.
}
The current $F_{zzz}$ is not only BRST closed, $j^{B}_{z}(z) F_{www}(w) = 0$, but even 
BRST exact 
\eqn\bbPhi{
j^{B}_{z}(z) \Phi_{www}(w) \sim {F_{www}(w) \over (z-w)}\,, ~~~ \Phi_{zzz} = -{i\over 2} \b^{m}_{z} \k_{z} \g_{m} \k_{z}\,.
}
The six  currents  $j^{B}_{z}, B_{zz}, j^{gh}_{z}, T_{zz}, F_{zzz}, \Phi_{zzz}$ 
generate a closed algebra which has the form of a Kazama algebra. This is expected:
quantization of gauged WZNW models generically leads to Kazama algebras instead of N=2 
superconformal algebras \Kazama~\getzler. 

In our earlier work with $b, c_z$ present, we only partially succeeded in constructing an operator $B_{zz}$ with 
the correct properties \Grassione, but the expression in \bigbb~satisfies all the desired properties.


\newsec{The gravitational topological Koszul quartet}

Consider a quartet $(b_{zz},c^{z},\beta_{zz}, \g^{z})$ containing the usual spin $(2,-1)$ 
gravitational ghosts, and spin $(2,-1)$ commuting counterparts $(\b_{zz}, \g^{z})$. The 
propagators are 
\eqn\gravA{
c^{z}(z) b_{ww}(w) \sim {1\over z-w}\,, ~~~~~~~~
\g^{z}(z) \b_{ww}(w) \sim {1\over z-w}\,.
}
From these fields we construct the energy-momentum tensor $T_{zz}$, the 
BRST current $j^{B}_{z}$, the ghost current $J^{gh}_{z}$, and an anticommuting 
spin 2 current $B_{zz}$
\eqn\gravB{
T_{zz} = - 2\b_{zz} \p_{z} \g^{z} - \p_{z} \b_{zz} \g^{z} - 2 b_{zz} \p_{z} c^{z} - \p_{z}b_{zz} c^{z}\,,
 }
$$j^{B}_{z} = - b_{zz} \g^{z}\,, ~~~~~~~~~~~ 
J^{gh}_{z} = - b_{zz} c^{z} - 2\b_{zz}\g^{z}\,,$$
$$B_{zz} = 2 \b_{zz} \p_{z} c^{z} + c^{z} \p_{z}\b_{zz} + \mu\, b_{zz}\,.$$
The real constant $\mu$ will be fixed later.
As usual $b_{zz}$ and $c^{z}$ have ghost number $-1$ and $+1$, respectively, but 
$\b_{zz}$ and $\g^{z}$ have ghost number $-2$ and $+2$, respectively. Hence, $B_{zz}$ has 
ghost number $-1$. 

The OPE's of these currents constitute a closed superconformal algebra
\eqn\gravC{
T_{zz}(z) T_{ww}(w) \sim { 2 T_{ww}(w) \over (z-w)^{2} } + {\p_{z} T_{zz} \over (z-w)}\,,
}
$$T_{zz}(z) j_{z}^{B}(w) \sim {j^{B}_{w}(w) \over (z-w)^{2} } + {\p_{z}j_{z}^{B} \over (z-w)}\,,$$
$$T_{zz}(z) B_{ww}(w) \sim { 2 B_{ww}(w) \over (z-w)^{2}} + {\p_{z} B_{zz} \over (z-w)}\,,$$
$$T_{zz}(z) J^{gh}_{w}(w) \sim {3 \over (z-w)^{3}} + {J^{gh}_{z} \over (z-w)^{2}} + 
{\p_{z} J_{z}^{gh} \over (z-w)}\,,$$
$$J_{z}^{gh}(z) J_{w}^{gh}(w) \sim {-3 \over (z-w)^{2}} \,,$$
$$j_{z}^{B}(z) B_{ww} \sim {-3 \over (z-w)^{3}} + {J_{z}^{gh} \over (z-w)^{2}} + 
{T_{ww} \over   (z-w)}  \,,$$
$$j^{B}_{z}(z) j^{B}_{w}(w) \sim 0\,, ~~~~~~~~~~~
    B_{zz}(z) B_{ww}(w) \sim 0\,,$$
 $$ 
 J^{gh}_{z} j^{B}_{w}(w)  \sim {j_{w}^{B}(w)  \over (z-w)}\,, ~~~~~~~~~~
 J^{gh}_{z} B_{ww}(w)  \sim {-B_{ww}(w)  \over (z-w)}\,,
 $$
The absence of an anomaly in the OPE of $T_{zz}$ with itself indicates that we are dealing with a 
twisted $N=2$ algebra. 
 It is clear that $T_{zz}, j_{z}^{B}$, and $B_{zz}$ are primary fields for any value of $\mu$, but the 
 ghost current $J^{gh}_{z}$ has an anomaly $+3$, which is opposite to the anomaly in $J_{z}^{gh}(z) J_{w}^{gh}(w)$ 
 and $j_{z}^{B}(z) B_{ww}(w)$. Furthermore, the BRST current and the $B$ field are nilpotent, while 
 $ j_{w}^{B} B_{zz}(z)$ reproduces $T_{zz}$ and $J_{z}^{gh}$. 
    
  We now add the currents $T^{W}_{zz},  
   j_{z}^{B, W}, B^{W}_{zz}, J^{gh, W}_{z}$ and $F^{W}_{zzz}, \Phi^{W}_{zzz}$ of the Kazama algebra 
   for the WZNW model to the 
   currents of the topological gravitational model. As shown in \figue, 
   one then ends up with an $N=2$ superconformal algebra, 
   provided one modifies the $B_{zz}$ field suitably. 
      The properly modified currents for the sum of both systems are given by
   \eqn\gravD{
\hat T_{zz} =   T^{W}_{zz} +   T^{top}_{zz} \,, ~~~~~~
\hat j_{z}^{B} = j^{B,W}_{z} +j^{B,top}_{z}\,,
   }  
$$
\hat J^{gh}_{z} = J^{gh, W}_{z} + J^{gh, top}_{z} \,,
$$    
$$
\hat B_{zz} = B^{W}_{zz} + B^{top}_{zz}(\mu =1) - {1\over 2} c^{z} F^{W}_{zzz} - {1\over 2}\g^{z} \Phi^{W}_{zzz}\,. 
$$  
 The currents of the Koszul quartet in \gravB\ 
 are denoted here  by the superscript $top$, while the currents of the $WZNW$ model 
 are denoted by the superscript $W$ and 
 can be found in eqs. (2.12), (4.1), (4.2), and (4.4). In particular 
 we recall the relations    
\eqn\gravE{
j_{z}^{B, W}(z) B^{W}_{ww} \sim {- 22 \over (z-w)^{3}} + {J_{z}^{gh,W} \over (z-w)^{2}} + 
{T^{W}_{ww} \over   (z-w)}  \,,
}   
$$B^{W}_{zz}(z) B^{W}_{ww}(w) \sim {F^{W}_{www}(w) \over (z-w)}\,,$$
$$j^{B}_{z}(z) \Phi^{W}_{www}(w) \sim {F^{W}_{zzz} \over z-w}\,, 
~~~~~~~~~
j_{z}^{B, W}(z) F^{W}_{www}(w) \sim 0\,.
$$
$$
B^{W}_{zz}(z) \Phi^{W}_{www}(w) \sim 0\,, ~~~~~~
B^{W}_{zz}(z) F^{W}_{www}(w) \sim {3 \Phi^{W}_{www}(w) \over (z-w)^{2}} + 
{\p_{w} \Phi^{W}_{www}(w) \over (z-w)}\,.
$$

It is now straightforward to verify that the currents in \gravD\ satisfy the same algebra as the currents in \gravB. 
For example, the conformal anomaly in the WZNW model cancels between the $g$-currents, the ghost-currents and 
the $h$-currents, while the Koszul quartet has no conformal anomaly, being topological. Furthermore, 
\eqn\gravF{
\hat T_{zz}(z) \hat J^{gh}_{w}(w) \sim {(22+3) \over (z-w)^{3}} + {\hat J^{gh}_{w}(w) \over (z-w)^{2}} + 
{\p_{w} \hat J_{w}^{gh}(w) \over (z-w)}\,,
}
$$
\hat J_{z}^{gh}(z) \hat J_{w}^{gh}(w) \sim {-22-3 \over (z-w)^{2}} \,,
$$
confirming that the algebra is twisted. Less obvious are the OPE's involving $\hat B_{zz}$, 
but they are of the form \gravB, too. For example, in
\eqn\gravG{
\hat j_{z}^{B}(z) \hat B_{ww}(w) \sim {- 22 - 3 \over (z-w)^{3}} + {\hat J_{z}^{gh}(w) \over (z-w)^{2}} + 
{\hat T_{ww}(w) \over   (z-w)}  \,,
}
the $\Phi^{W}_{zzz}(z)$ and $F^{W}_{zzz}(z)$ terms cancel. The most interesting case is $\hat B_{zz}(z) \hat B_{ww}(w)$ which should vanish and does vanish. 
Another good check on the $F^{W}_{zzz}$ and $\Phi^{W}_{zzz}$ terms in $\hat B_{zz}$ 
is $\hat j^{B}_{z}(z) \hat B_{ww}(w)$ which should be independent of $F^{W}_{zzz}$ and 
$\Phi^{W}_{zzz}$; this is indeed the case. 


\newsec{Definition of Physical States} 

Having constructed the BRST charge $Q_{W}$ according to the quantization prescription 
for WZNW models, we must now define the physical states. It is easy to see that the 
cohomology of $Q_{W}$ by itself does not yield the correct spectrum for the superstring. As we 
shall discuss in more detail below, 
the field equations for the cohomology depend on  
the coordinates $x + x^{h}$, 
$\t + \t^{h}$ and $p_{z\a} + p^{h}_{z\a}$, 
while the dependence on the coordinates $x-x^{h}$, $\t - \t^{h}$ and 
$p_{z\a} - p^{h}_{z\a}$ is not fixed. 
Thus, we have to follow a  different path. 

In purely topological models, there exists a second BRST charge, $Q_{V}$ given in the introduction, 
namely 
\eqn\bbrrssttA{
Q_{V} = \oint c^{z} \Big(T^{W}_{zz} + {1\over 2} T^{top}_{zz}\Big) + \g^{z} \Big(B^{W}_{zz} + 
{1\over 2} B^{top}_{zz} \Big)\,.
}
where the currents $T^{W}_{zz}, T^{top}_{zz}, B^{W}_{zz}$ and  $B^{top}_{zz}$ are given in the previous 
section. As can be easily checked the above BRST charge anticommutes with the BRST charge 
\eqn\bbrrssttB{
Q_{S} = Q_{W} + \oint \eta_{z} e^{\phi} b_{zz} + \oint \eta_{z}\,,
}
where the bosonic ghosts $\gamma^{z}$ and $\beta_{zz}$ are fermionized 
in the usual way, namely $\gamma^{z} = \eta_{z} e^{\phi}$ and $\beta_{zz} = \p_{z} \xi e^{-\phi}$ with 
$\xi(z) \eta_{w}(w) \sim (z-w)^{-1}$ and $\phi(z) \phi(w) \sim - \ln(z-w)$  
(see for example \lref\disl{
J.~Distler,
{\it 2-D Quantum Gravity, Topological Field 
Theory And The Multicritical Matrix Models,}
Nucl.\ Phys.\ B {\bf 342}, 523 (1990)} \disl).  The operator $\oint \eta_{z}$ is added 
for the same reasons as in the RNS formalism working in the large Hilbert space 
containing the zero mode $\xi_{0}$. The BRST charge is the 
sum of the BRST charges in the matter sector and in the topological sector, 
\eqn\bigQ{
{\bf Q} = Q_{S} + Q_{V}\,,
}
The physical states are therefore identified by 
the BRST cohomology ${\bf Q}$ and by the grading condition 
formulated in \Grassitwo. Neglecting the topological gravity sector, 
it is easy to show that one recovers the correct equations of motion for 
the massless sector of open and closed string theory. (Essentially, 
the formulas for the equations of motion are given in \Grassitwo\ and  
they are related to the present ones by a similarity transformation on the 
superfields,  $\tilde{A}_{\a} = e^{\cal R} A_{\a}$, etc... where 
${\cal R} = \t^{\a} D^{h}_{\a} + (x^{m} - i \t \g^{m}_{\a\b} \t^{h}) \p^{h}_{m}$). 
The complete analysis of the cohomology in the topological gravity sector and in 
 the matter sector follows the description given by \dvv\ and it will be given a 
 separate publication. 

The physical states are obviously defined up to gauge transformations, given 
by exact BRST vertex operators. This gauge freedom allows us to impose 
some gauge fixing conditions to remove the redundancy in the definition of 
physical states. Two important examples in the literature are the bosonic 
string and the fermionic string in the RNS formulation. In the bosonic string 
one imposes the Siegel condition: $b_{0} | {\rm phys} \rangle = 0$.  In the 
RNS string one imposes the superpartner condition $\b_{0} | {\rm phys} \rangle = 0$. 
In the RNS string one has also to remember that in order to have a non-trivial 
cohomology one has to require that vertices are polynomials in the 
zero mode of the superghost $\gamma_{0}$ \big. 

In our formalism, in order to gauge completely the model, we introduced in the previous section 
a replica of our the coordinates $x^{m}, \t^{\a}$ and $p_{z\a}$. The physical 
states will therefore depend on the combination not annihilated by the BRST charge 
$Q_{S}$. In order to remove this redundancy one has to  impose 
a gauge fixing condition and following the suggestion of the bosonic string, 
we impose the condition $B_{0}  | {\rm phys} \rangle = 0$ where 
$B_{0} = \oint z B_{zz}$. The requirement $B_{0}|{\rm phys} \rangle =0$ 
is imposed by hand at this point, and one also expects that one should impose 
$b_{0} |{\rm phys} \rangle =0$ and $\beta_{0} |{\rm phys} \rangle =0$ as in the 
RNS framework. Imposing all three conditions seems too much. Fortunately, as we now show, 
one of these conditions follows from the others. 

First of all, we observe that 
\eqn\bnot{
 \Big\{b_{0}, Q_{S} + Q_{V} +\oint \eta_{z}\Big\} = \hat L_{0} \,,
}
where $\hat L_{0} = \oint z (T^{W}_{zz} + T^{top}_{zz})$ is the Virasoro generator of the 
combined system. Therefore, imposing the gauge fixing condition $b_{0} | {\rm phys} \rangle = 0$ 
we obtain the usual constraint on the physical states $\hat L_{0} | {\rm phys} \rangle = 0$. The 
gauge fixing might be formulated in string field theory context as the Siegel gauge. 

On the other hand, we can also fix the gauge symmetry (notice that one gauge invariance is 
generated by $Q_{V}$ and the other by $Q_{S}$) by the gauge choice 
$\beta_{0}| {\rm phys} \rangle =0$. 
Using the fact that 
\eqn\bbnot{
\Big[ \beta_{0}, Q_{S} +Q_{V} + \oint dz \eta_{z} \Big] = b_{0} + B_{0}\,, 
}
we finally deduce the condition $B_{0}|{\rm phys} \rangle =0$. The latter condition removes the dependence 
on the combinations $J^{g}_{M} - J^{h}_{M}$, not fixed by the BRST charge $Q_{W}$. 

The result of our analysis is the definition of physical states given in the introduction. 
 
 \newsec{Equivalence with our former approach}

The present new formulation of the superstring in terms of a WZNW model 
prompts us to ask how it is related to our earlier work \grassix\ with the $b,c_z$ multiplet. 
Consider the composite operator which contains the $h$-currents in (2.12)
\eqn\caz{
{\cal J}_z = -  \l^\a J^h_{z\a} - \xi^m J^h_{z m} - \chi_\a J^{h,\a}_{z}\,.}
Due to the statistics of the ghosts $\xi^m, \l^\a, \chi_\a$, the OPE of two of these currents 
contains only first-order poles
 \eqn\cicd{
 {\cal J}_z(z) {\cal J}_w(w) \sim { - \xi^m \p_w \xi_m + i \chi_\a \p_w \l^\a - i\l^\a \p_w \chi_\a 
 - 2 i (\l \g^{m} \l) J^{h}_{m} +  4 \xi^{m} (\l \g_{m} J^{h}) \over z-w} \,.
 }
 Furthermore, the new current 
 \eqn\newcic{
 {\cal S}_z = \xi^m \p_z \xi_m -i  \chi_\a \p_z \l^\a + i \l^\a \p_z \chi_\a 
 + 2 i (\l \g^{m} \l) J^{h}_{z m} -  4 \xi^{m} (\l \g_{m} J^{h}_{z})
 }
 has no OPE's with itself but with ${\cal J}_z(z)$ it yields 
 \eqn\newcicB{
 {\cal J}_{z}(z) {\cal S}_{w}(w) \sim { 3 {\cal P}(w) \over (z-w)^{2} } + { \p_{w} {\cal P}(w) \over (z-w) } \,,
  }
  with ${\cal P}(z) = 2 i\, \xi^{m} \l \g_{m} \l$. 

We first construct a new BRST operator $Q'$, starting from the generators $({\cal J}_z, {\cal S}_z, {\cal P})$ 
and  adding new ghosts $(\gamma, b, \tilde\b_{z})$ 
whose antighosts are $(\beta_{z}, c_z, \tilde\g)$
 \eqn\cice{
 Q' = \oint dz \Big( \gamma {\cal J}_z + {1\over 2} b {\cal S}_z + \tilde\b_{z} {\cal P} + \gamma^2 c_z
  + y \g b \p_{z} \tilde\gamma + z \g \p_{z} b \tilde\gamma + t \p_{z}\g b \tilde\g \Big) \,.
}
 The last four terms are the usual ghost-ghost-antighost term in the BRST charge. 
 For $y=1/2, z=1$, and $t= -1/2$ the BRST charge $Q'$ is nilpotent. Since $Q'$ does not 
 contain the antighost $\b_{z}$, and since the ghost number of $\gamma$ as well as its conformal 
 spin vanishes and, finally, since $\gamma$ is a commuting field, it can be set to unity. 

  Consider now the BRST charge $Q_{W}$ in (2.12). It can be rewritten by adding to the terms with 
  $g$-currents and ghost-currents the terms that yield the BRST charge $Q_{old}$ of our earlier work \grassix, while 
  the terms with the $h$-currents can be extended to yield $Q'$ 
  \eqn\cicf{
  Q_{W} = Q_{old} + Q' + \oint dz \Big[ - 2 c_{z} - {\rm terms~with}~\tilde\beta_{z} ~{\rm and~} \tilde\gamma\Big]
  }
  $$
  Q_{old}= \oint dz \Big[ j_{z}^{B}|_{J^{h} =0} + c_{z} - {1\over 2} b {\cal S}_{z}\Big]
  $$
Because the double poles in the OPE's of the $h$-currents differ by a sign from the corresponding poles 
of the $g$-currents, the anomaly (proportional to ${\cal S}_{z}$) in $Q_{old}$ has the opposite sign to the anomaly 
in $Q'$. 

Our original BRST charge can thus be written as 
\eqn\cicg{
Q_{old} = Q_{W} - Q' + 2 \oint c_{z} + \oint ({\rm ~terms ~with}~ \tilde\beta_{z}~{\rm and}~\tilde\gamma) \,.
}
It contains the difference of the two nilpotent charges $Q_{W}$ and $Q'$. The anticommutator of $Q_{W}$ 
and $Q'$ is cancelled by the anticommutator of $2 \oint c_{z}$ with $Q'$, and the contributions with $\tilde\beta_{z}$ and 
$\tilde\gamma$ cancel in the square of the right-hand side. The remaining charge, $Q_{old}$ is thus nilpotent, 
as indeed it was found to be. 

We see that the ghost pair $b, c_z$ has the surprising 
interpretation that $b$ is actually a ghost and  $c_z$ is an antighost. This explains why $c_z$ has conformal spin 1 as 
all the other antighosts. 
 

\newsec{Conclusions and Outlook}

We have obtained a covariant quantum superstring with 
manifest spacetime Lorentz invariance. It is based on a WZNW 
model, which itself is based on a particular non-semisimple 
Lie superalgebra, namely the super-Poincar\'e algebra with 
a fermionic central extension. No ghosts were any longer added 
by hand as in our earlier work; rather, the ghost structure directly 
follows from the requirement that our theory is invariant under 
superdiffeomorphisms. For this reason we added a quartet of ghosts 
$(b_{zz}, c^{z}, \beta_{zz}, \g^{z})$ which is needed for 
the gravitational sector. The currents for the combined system 
satisfy an ordinary $N=2$ superconformal algebra, which 
raises the hope that quantum computation may be easier 
that originally thought, and that the classical geometrical 
meaning of $\kappa$-symmetry may become clear. 

Physical states are defined as follows: they lie in the cohomology 
$Q_{S} + Q_{V}$ defined in section 6, have vanishing grading, 
and are annihilated by $b_{0}$ and $\beta_{0}$. The latter two conditions 
are gauge choices and select particular representatives of the 
cohomology; they are the superextension of the well-known 
Siegel gauge $b_{0} | {\rm phys }\rangle= 0$ of the bosonic string. 
The deeper meaning of the grading condition still eludes us, 
but we are studying this problem. The need for a grading 
condition is, as in our earlier work, that one obtains a nontrivial 
cohomology. The ghosts $\beta_{zz}$ and $\gamma^{z}$ are fermionized 
into the fields $\xi, \eta_{z}$ and $\phi$, with $\xi$ having grading 2, 
just like the field $b$ in our earlier work, and $\eta_{z}$ having grading 
$-2$. 
A complete discussion of the cohomology will be published 
elsewhere, but it is clear that it will contain the cohomology 
we obtained in our earlier work for the matter sector, and the 
cohomology of the gravitational sector as already discussed by many 
authors. 


\vskip .5cm
\noindent
{\bf Acknowledgements}~
\vskip .5cm

\noindent
At the 2002 Amsterdam Summer 
String Workshop E. Verlinde suggested to go on with adding ghosts in order to 
obtain BRST nilpotency, instead of adding by hand the $b, c_z$ system. 
At the 2003 Amsterdam Summer String Workshop H. Verlinde suggested to introduce 
the topological gravity quartet. This article contains the result of their suggestions. 

We thank the Ecole Normale Sup\'erieure at Paris and its 
director Eug\`ene Cremmer for a stay of a month during which this work was begun. 
Part of this research was supported by NSF grant PHY-0098527. GP is supported by PPARC. 



\newsec{Appendix A: Massless Vertex for Open Superstrings with $\phi_{\a}$} 

One can in principle work with the field $\phi_{\a}$ present in the theory. In this appendix, we discuss how 
to recover the correct cohomology and, therefore, the correct spectrum of the theory in presence 
of the field $\phi_{\a}$. We follow technique of our previous derivations 
\grassix. Instead using the auxiliary currents, we obtain a straigthforward derivation of the massless spectrum of 
by using the $b,c_z$ system and the grading. 
From the construction of the previous appendix, we use the derivatives $D_{\a}, \p_{m}$ 
and $D^{\a}$ which satisfies the commutation relations 
\eqn\dercommu{
\{ D_{\a}, D_{\b} \} = - 2 i \g^{m}_{\a\b} \p_{m}\,, ~~~~~~
[ D_{\a}, \p_{m} ] = - 2 \g^{m}_{\a\b} D^{\b}\,, ~~~~~~
\{ D_{\a}, D^{\b} \} = 0\,,
}
$$
[\p_{m}, \p_{n}] = 0\,, ~~~~~
[\p_{m}, D^{\a}] = 0\,, ~~~~~
[D^{\a}, D^{\b} ] = 0\,.
$$
which depend on the field $\phi_{\a}$. In addition, we define the usual derivatives 
$D^{o}_{\a}$ and $\p^{o}_{m}$, independent of $\phi_{\a}$ which satisfy the 
usual superspace relations $\{ D^{o}_{\a}, D^{o}_{\b} \} = 
- 2 i \g^{m}_{\a\b} \p^{o}_{m}$ and $[ D^{o}_{\a}, \p^{o}_{m} ] = 0$. 

If we keep $\phi_{\a}$ in the theory, the superfields $A_{\a},  \dots , F^{\a\b}$ in the vertex operator 
depends on the coordinate $x^{m}, \t^{\a},\phi_{\a}$ and the corresponding $h$-partners. 
The vertex ${\cal U}^{(1|0)}(z)$ belongs to the space of zero graded polynomials 
(following the grading assignment given in \Grassitwo) and the cohomology is defined by 
\eqn\cohoD{
\{ {\bf Q}, {\bf U}^{(1|0)}(z) \} = 0\,, ~~~~~~~ \delta {\bf U}^{(1|0)} = [{\bf Q}, \Omega(z) ]\,,
}
 where the gauge parameter superfields $\Omega$ is a function of $x^{m}, \t^{\a}$ and $\phi_{\a}$. 
 Computing \cohoD, we obtain the following equations (we neglect the contributions of the 
 $\omega$-dependent terms since they are cohomologically trivial anyway). 
 The condition $\{ {\bf Q}, {\cal U}^{(1|0)}(z) \} = 0$ implies the following equations 
\eqn\cohoE{\eqalign{ 
& D_{(\a} A_{\b)} - \half \g^m_{\a\b} A_m = 0 \,, \cr 
& \p_m A_\a - D_\a A_m + \g_{m\a\b} W^\b = 0 \,, \cr 
& \p_{[m} A_{n]} + F_{mn} = 0 \,,  \quad\quad 
D_\b W^\a +   F_\b^{~~\a} = - D^{\a} A_{\b}  \,, \cr
& \p_m W^\a + F_{~~m}^{\a} = D^{\a} A_{m}  \,, \quad\quad 
F^{\a\b}  =  - D^{(\a} W^{\b)}  \,, \cr
}}
where the terms on the right side are due to the $\phi_{\a}$-dependence of 
${\bf U}^{(1|0)}$. These field equations are invariant under the transformations 
$ \delta {\bf U}^{(1|0)} = [{\bf Q}, \Omega(z) ]$
\eqn\cohoF{
\delta A_{\a} = D_{\a} \Omega\,, ~~~~~
\delta A_{m} = \p_{m} \Omega\,, ~~~~~
\delta W^{\a} = D^{\b} \Omega\,,
}
$$
\delta F_{mn} =0\,, ~~~~~
\delta F_{\a}^{~\b} = 0\,, ~~~~
\delta F_{m}^{~\b} = 0\,, ~~~~
\delta F^{\a\b} = 0\,.
$$
In order to remove the field dependence $\phi_{\a}$, we have to impose 
an extra condition. This can be done easily by using the charge $K^{\a} = \oint i \p_{z} \t^{\a}$  
we discussed in the introduction. The charge $K^{\a}$ commute with the BRST charge, it 
is an anticommuting  nilpotent operator, and the physical states are defined by 
\eqn\phicoho{
\{ {\bf Q}, \, U \} = 0\,, ~~~~~ \{ K^{\a}, U\} = 0\,, ~~~~ \delta U = [ Q, \Omega ]\,, ~~~~
[K^{\a} , \Omega] =0\,,
}
The physical states are defined as the equivariant cohomology of $Q$ with respect 
to the gauge transformations generated by $K^{\a}$ which correspond to constant shifts of the 
field $\phi_{\a}$.  This allows us to remove the zero modes of $\phi_{\a}$ from the theory and 
the vertex operators will depend only on derivatives of $\phi_{\a}$. 

The conditions $\{ K^{\a}, U\} = 0$ and $[K^{\a} , \Omega] =0$ imply that 
$D^{\a} A_{\b} =0, \dots, D^{\a} F^{\b\g} =0$ and $D^{\a} \Omega =0$ removes the 
dependence on the zero mode of $\phi_{\a}$. Therefore, all derivatives in eqs.~\cohoE~become 
the usual derivatives $D^{o}_{\a}$ and $\p^{o}_{m}$. Moreover, the resulting field equations 
coincide with the usual equation obtained in \berkox~and \grassix. 
  
      
\newsec{Appendix B: The Relation between $\phi_{\a}$ and $p_{z \a}$}

One can clarify the relation between $\phi_{\a}$ and $p_{z \a}$ by evaluating how for example $\t^{\a}$ and $x^{m}$ 
transform under susy. The susy generator can be written in two different ways 
\eqn\susyA{
Q^{susy}_{\a} = 
\oint dz (i \p_{z}\phi_{\a}) = 
\oint \Big[ p_{z\a} - i \p_{z} x^{m} (\g_{m} \t)_{\a} - {1\over 6} (\g^{m} \t)_{\a} (\t \g_{m} \p_{z} \t) \Big]\,.
}
From the right-hand side one obtains straigthforwardly $\delta \t^{\a} = 
[ \e^{\b} Q^{susy}_{\b}, \t^{\a}] = \e^{\a}$ and $\delta x^{m} = i \e \g^{m}\t$, which of course leave $\Pi^{m}_{z}$ invariant. 
To obtain the same results from the left-hand side we use the WZNW action \WZW~and apply perturbation 
theory in the interaction picture. The action is the $\int d^{2}u$ integral of 
\eqn\WZNWexp{
{\cal L}_{g}(u) = {1\over 2} \Pi^m_z \Pi^{n}_{ \z} \eta_{mn} + d_{z\a} \bar\p \t^\a + \hat d_{\z \a} \p \hat\t^{\a} = 
}
$$
{1\over 2} 
\Big( \p_{z} x_{m} - i \t \g_{m} \p_{z}\t - i \hat\t \g_{m} \p_{z} \hat\t \Big)
\Big( \bar\p_{\z} x^{m} - i \t \g^{m} \bar\p_{\z}\t - i \hat\t \g^{m} \bar\p_{\z} \hat\t \Big) + 
$$
$$
i\, \Big( \p_{z} \phi_{\a} +  2 \, \p_{z} x^{m} (\g_{m} \t)_{\a} -  
{  2 i \over 3} (\g_{m} \t)_{\a} (\t \g^{m} \p_{z} \t)\Big) \bar\p_{\z} \t^{\a}+ 
$$
$$
i\, \Big(\bar\p_{\z} \hat\phi_{\a} +  2 \, \bar\p_{\z}x^{m} (\g_{m} \hat\t)_{\a} -  {  2 i \over 3} (\g_{m} \hat\t)_{\a} 
(\hat\t \g^{m} \bar\p_{\z} \hat\t ) \Big) \p_{z} \hat\t^{\a} \,. 
$$
The $\phi$-$\t$ propagator reads 
$\phi_{\a}(z,\z) \t^{\b}(w,\bar w) \sim - i \ln | z-w |^{2} \delta_{\a}^{\b}$, 
and using it in \susyA~reproduces $\delta \t^{\a} = \e^{\a}$, 
but for $\delta x^{m}$ we need interaction vertices. The vertices with $\hat\t^{\a}$ cannot contract with $\phi_{\a}$ 
hence we need only the vertices $x \t\t$. This yields  
\eqn\qx{
[Q^{susy}_{\a} , x^{m}] = \Big[ \oint_{\g_{w}} dz \,  (i \p_{z} \phi_{\a}) \Big] 
\Big[ \int d^{2}u \Big( - {i \over 2} \p x^{m} (\t \g_{m} \bar\p \t)  - {i\over 2} \bar\p x^{m} (\t \g_{m}\p \t) \Big) \Big] 
\Big[ x^{m}(w) \Big]\,,
}
where the $z$-integral is taken along the curve $\g_{w}$ which encircles $w$. 
The vertices with $\bar\p x^{m}$ yield total $u$-derivatives and integration over $d^{2}u$ yields a 
vanishing result. However, the vertices with $\bar\p x^{m}$ contribute because
\eqn\qy{
i \p \phi_{\a}(z) \,  \bar\p \t^{\b}(u) = \bar\p_{\bar u} \Big({1\over z-u}\Big) = - \pi \delta^{2}(z-u)\,.
}
Both $\t$'s in $\p x^{m} \t \g_{m} \bar\p \t$ contribute (one needs one partial integration) and after ordinary integration 
over $u$ and contour integration of $z$, one obtains the correct result. At tree graph level, diagrams with more 
than one interaction vertex do not contribute because they are disconnected graphs.


\listrefs  
  
\bye